\documentclass{natureprintstyle}
\usepackage{graphicx}
\usepackage{amssymb}
\usepackage[T1]{fontenc}
\usepackage[lowtilde]{url}
\usepackage{hyperref}
\usepackage{newfloat}
\DeclareFloatingEnvironment[name={Extended Data Figure}]{extfig}

\newcommand{\kms}{\hbox{km s$^{-1}$}}
\newcommand{\re}{\hbox{${\rm R}_{\rm e}$}}

\newcommand{\mbh}{$M_\bullet$}
\newcommand{\msun}{\hbox{M$_{\odot}$}}

\title{Black-hole regulated star formation in massive galaxies}

\author{Ignacio Mart\'in-Navarro$^{1,2}$,
Jean P. Brodie$^{1}$,
Aaron J. Romanowsky$^{3,1}$,
Tom\'as Ruiz-Lara$^{4,5}$
\& Glenn van de Ven$^{2,6}$
}

\begin{document}

\maketitle

\let\thefootnote\relax\footnote{
\begin{affiliations}
\item University of California Observatories, 1156 High Street, Santa Cruz, CA 95064, USA
\item Max-Planck Institut f\"ur Astronomie, Konigstuhl 17, D-69117 Heidelberg, Germany
\item Department of Physics and Astronomy, San Jos\'e State University, One Washington Square, San Jose, CA 95192, USA
\item Instituto de Astrof\'isica de Canarias, E-38200 La Laguna, Tenerife, Spain
\item Departamento de Astrof\'isica, Universidad de La Laguna, E-38205 La Laguna, Tenerife, Spain
\item European Southern Observatory, Karl-Schwarzschild-Str. 2, 85748 Garching b. M\"unchen, Germany
\end{affiliations}
}

\begin{abstract}

Super-massive black holes, with masses larger than a million times that of the Sun, appear to inhabit the centers of all massive galaxies\cite{Magorrian98,Gebhardt00}. Cosmologically-motivated theories of galaxy formation need feedback from these super-massive black holes to regulate star formation\cite{Silk12}. In the absence of such feedback, state-of-the-art numerical simulations dramatically fail to reproduce the number density and properties of massive galaxies in the local Universe\cite{Vogelsberger14,Schaye15,Choi17}. However, there is no observational evidence of this strongly coupled co-evolution between super-massive black holes and star formation, impeding our understanding of baryonic processes within galaxies. Here we show  that the star formation histories (SFHs) of nearby massive galaxies, as measured from their integrated optical spectra, depend on the mass of the central super-massive black hole. Our results suggest that black hole mass growth scales with gas cooling rate in the early Universe. The subsequent quenching of star formation takes place earlier and  more efficiently in galaxies hosting more massive central black holes. The observed relation between black hole mass and star formation efficiency applies to all generations of stars formed throughout a galaxy's life, revealing a continuous interplay between black hole activity and baryon cooling. 

\end{abstract}

\begin{figure}[h]
\begin{center}
\includegraphics[width=8.5cm]{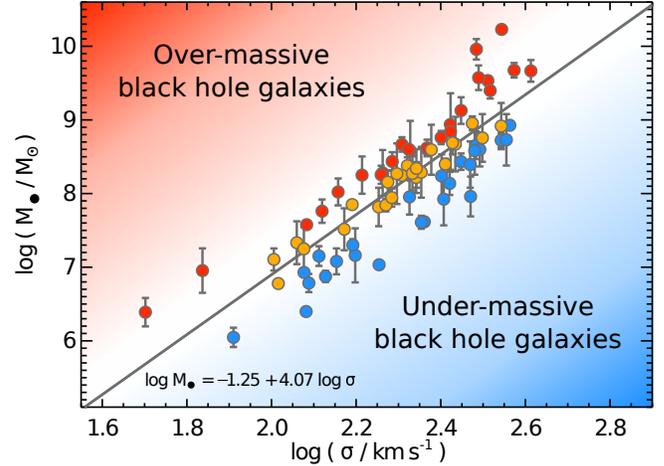} 
\end{center}
\caption{{\bf Black hole mass -- stellar velocity dispersion relation.} The stellar velocity dispersion of galaxies ($\sigma$) tightly correlates with the mass of their central super-massive black hole (\mbh).  Data points correspond to the 74 HETMGS galaxies with measured black hole masses and high-quality spectra. The solid line indicates the average black hole mass for a given velocity dispersion. Galaxies above this best-fitting \mbh--$\sigma$ relation have black holes more massive than expected for their velocity dispersion, and therefore are called over-massive black hole galaxies (red). Conversely, galaxies hosting lighter central black holes than the average population (blue) are called under-massive black hole galaxies. Galaxies with standard black hole masses are shown in orange.}
\label{fig:relation}
\end{figure}

As shown in Figure~\ref{fig:relation}, the mass of super-massive black holes (\mbh) scales with the stellar velocity dispersion ($\sigma$) of their host galaxies\cite{Ferrarese00,Gebhardt00}. The scatter in this relation can be used to quantify how massive a given black hole is compared to the average population. We can then define over-massive and under-massive black hole galaxies as those objects lying above and below the best-fitting \mbh-- $\sigma$ relation, respectively. In other words, over-massive black hole galaxies have central black holes more massive than expected, whereas under-massive black hole galaxies host relatively light super-massive black holes. The distinction between these two types of galaxies allows us to evaluate the role of black hole activity in star formation, as the amount of energy released into a galaxy is proportional to the mass of the black hole\cite{Crain15,Sijacki15}.

We based our stellar population analysis on long-slit optical spectra from the Hobby-Eberly Telescope Massive Galaxy Survey\cite{hetmgs}. The resolution of the data varies between 4.8 and 7.5 \AA, depending on the slit width. We adopted a fixed aperture of half the effective radius 0.5\,\re, where \re \ is defined as the galactocentric radius that encloses half of the total light of a galaxy. This aperture is large enough to allow for a direct comparison in the future between our results and numerical simulations, but also small enough to ensure that we are dominated by {\it in situ} star formation\cite{RG16}. The sizes of all galaxies in our sample were calculated in a homogeneous way using infrared $K$-band photometry\cite{vdb16}. We focused on spectroscopic analysis of wavelengths between 460 and 550 nm, covering the most prominent age and metallicity spectral features in the optical range.

SFHs were measured using the STEllar Content and Kinematics via Maximum A Posteriori likelihood (STECKMAP) code\cite{Ocvirk06b}, fed with the MILES stellar population synthesis models\cite{miles}. STECKMAP is a Bayesian method which decomposes the observed spectrum of a galaxy as a temporal series of single stellar population models. Its ability to recover reliable SFHs of unresolved systems has been thoroughly tested\cite{Koleva08,Pat11,Leitner12}. Furthermore, STECKMAP-based SFHs are in remarkable agreement with those based on color--magnitude diagrams of nearby resolved systems\cite{rl15}. Our SFHs are reliable in a relative sense (see Methods) even if there are 
systematics from the limited set of models.

Our final sample consists of all HETMGS galaxies with direct black hole mass measurements, and for which we can also determine their SFHs, 74 in total, probing total stellar masses from M$\sim1\times10^{10}$\msun \ to M$\sim2\times10^{12}$\msun. We removed from the final sample galaxies with strong nebular emission lines, in particular around the optical H$\beta$ line, which affected the quality of the STECKMAP fit. Note that galaxies with significant emission lines populate only the low-$\sigma$ end of our sample ($\log \sigma \lesssim 2$). We normalized individual SFHs so that each galaxy has formed one mass unit at $z\sim0$. 

\begin{figure*}
\begin{center}
\includegraphics[width=9.1cm]{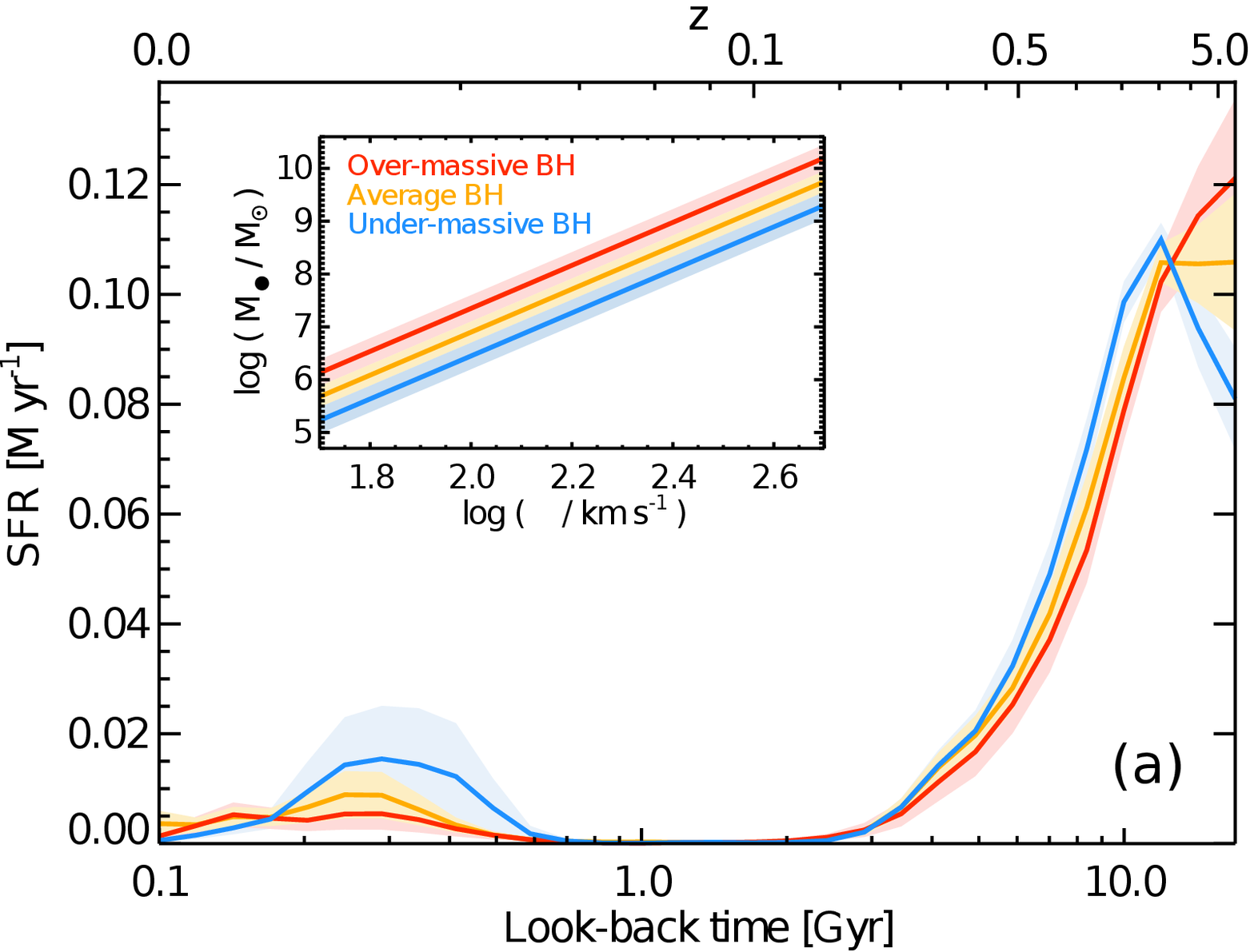}
\includegraphics[width=8.9cm]{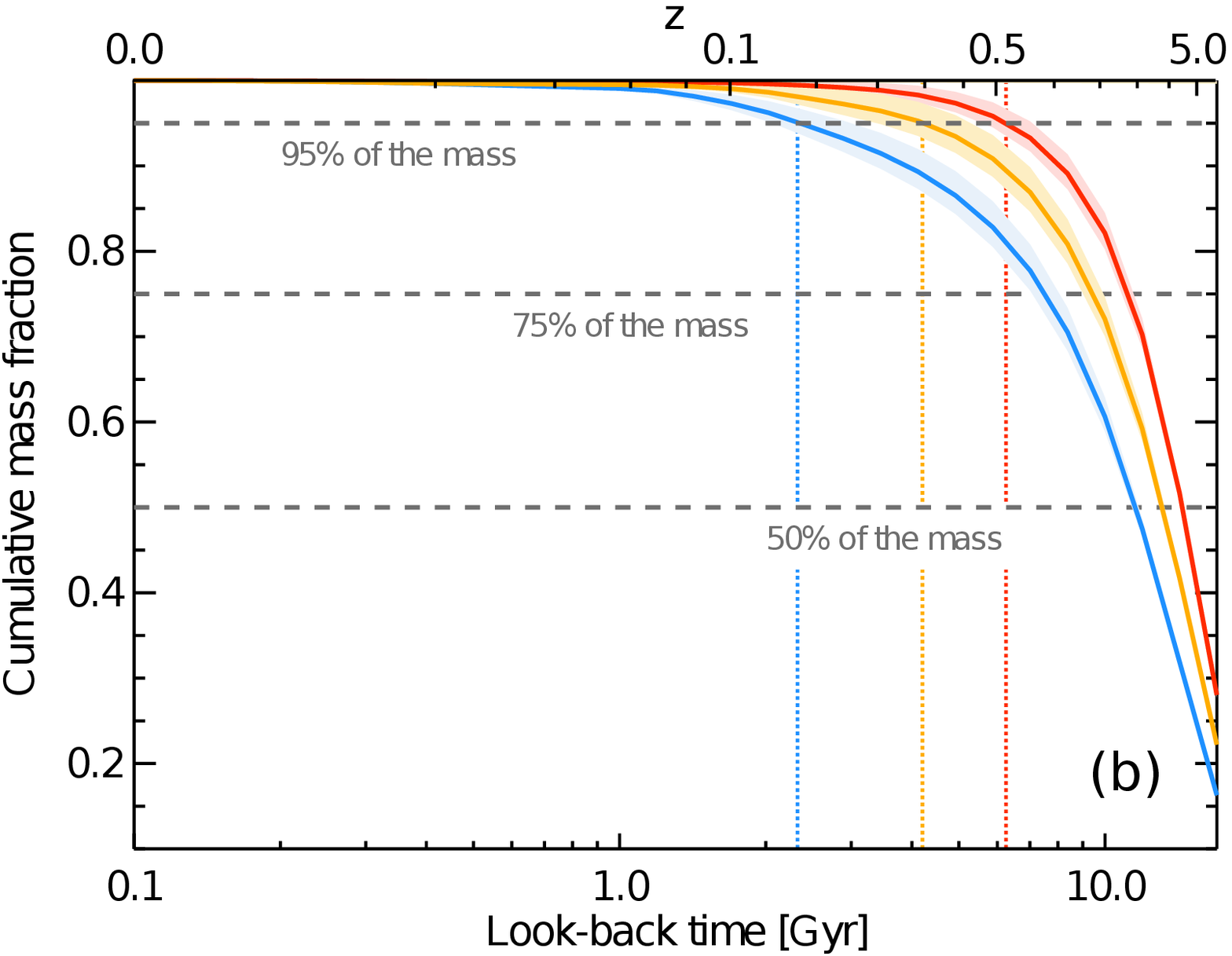}
\end{center}
\caption{{\bf Star formation evolution over cosmic time.} Red, orange, and blue solid lines correspond to over-massive, standard, and under-massive black hole galaxies, respectively. The light shaded regions indicate the 1-$\sigma$ uncertainties in the mean values (but not model uncertainties). Panel (a) shows the evolution of the star formation rate (SFR) as a function of look-back time and panel (b) shows the cumulative mass distribution for the three types of galaxies. Vertical dotted lines in (b) indicate when 95\% of the final mass has been reached. The coupling between black hole mass and star formation applies to all generations of stars, from those formed at $z\sim5$, to the youngest generations. Our measurements of the star formation history as a function of black hole mass show how the latter drives baryonic cooling within massive halos.}
\label{fig:sfr}
\end{figure*}

Our main result is summarized in Figure~\ref{fig:sfr}, where we show how SFRs and cumulative stellar mass have evolved in over-massive (red), in standard (orange) and in under-massive (blue) black hole galaxies. This evolution of star formation over cosmic time is strongly coupled to the mass of the central black hole. Over-massive black hole galaxies, i.e., those with relatively more massive black holes, experienced more intense SFRs in the very early Universe (look-back times $\gtrsim 10$ Gyr) compared to galaxies with lighter black holes.  Star formation in over-massive black hole galaxies was quenched earlier, reaching 95\% of their final mass, on average, $\sim4$~Gyr earlier than under-massive black hole galaxies, as shown by the cumulative mass distributions. The amount of recent star formation, however, inversely correlates with the mass of the black holes, i.e., young stellar populations are more prominent in under-massive black hole galaxies. A Kolmogorov-Smirnov test of the distributions shown in Figure~\ref{fig:sfr} indicate that they are significantly different (P-value = 0.026).

It is worth emphasizing that SFHs and black hole masses are completely independent observables. Black hole masses were calculated using a wide variety of methods, but without detailed information on the stellar population properties\cite{vdb16}. If the differences presented in Figure~\ref{fig:sfr} were to be artifacts of the stellar populations analysis, they could not be coupled to the mass of the black hole. This relative character of our approach minimizes the effect of systematic errors in the analysis. We have further checked that neither our choice for the STECKMAP free parameters, nor the adopted \mbh--$\sigma$ relation, nor the stellar populations modeling affects our conclusions (see Methods). Note also that there is no significant difference in the velocity dispersion of under-massive, standard and over-massive black hole galaxies ($\log \sigma_\mathrm{under}=2.32 \pm 0.04$, $\log \sigma_\mathrm{standard}=2.30 \pm 0.03$, $\log \sigma_\mathrm{over}=2.32 \pm0.04$). 

It could be argued that the process regulating black hole growth also affected the efficiency of baryonic cooling within galaxies. In particular, objects formed in high density environments could have grown more massive black holes and formed their stellar populations differently due to the amount and properties of the available gas. However, the lack of a morphological offset across the \mbh-$\sigma$ relation\cite{Beifiori12,vdb16}  disfavors such a scenario (see also Methods). In addition, differences in formation timescales such as those shown in Figure~\ref{fig:sfr} do not depend on galaxy environment\cite{Thomas05}. Thus, over- and under-massive black hole galaxies have likely experienced similar formation paths. It is worth noting here the relative character of our analysis, i.e., independent of the normalization of the \mbh-$\sigma$ relation. Moreover, the robustness of our results with respect to additional parameters such as galaxy size or stellar density (see Methods) further suggest that over-massive and under-massive black hole galaxies are only different in terms of their detailed stellar population properties and black hole masses. Dynamically, morphologically and structurally, both types of galaxies are indistinguishable.

The measurements shown in Figure~\ref{fig:sfr} probe the star formation processes within massive halos since the early Universe. Interestingly, black hole masses and star formation seem to be related as early as $z\sim5$. This invalidates any scenario where the observed scaling relations between black holes and host galaxies would emerge non-causally from the hierarchical evolution of a $\Lambda$-CDM Universe\cite{Peng07,Jahnke11}. At the peak of the SFR, baryon cooling was more efficient in galaxies with (present-day) more massive black holes. The stellar mass formed around $z\sim5$ in over-massive black hole galaxies is $\sim 1.3$ times that formed within under-massive black hole galaxies. Assuming the stellar-to-black hole mass ratio observed in the local Universe\cite{vdb16}, these differences in the amount of stellar mass formed at $z\sim5$ imply that more than 50\% of the (vertical) scatter in the \mbh--$\sigma$ relation results from this initial phase of galaxy formation and black hole growth. We hypothesize that over-massive black hole galaxies rapidly reached a black hole mass capable of quenching star formation, which led to a shorter star formation time-scale. Thus, the baryon cooling efficiency at high redshift would play a major role in determining the present-day mass of super-massive black holes, feeding the primordial black hole seeds with gas, in agreement with quasar observations\cite{Fan01}.

The importance of super-massive black holes in galaxy evolution arises from their potential role as quenching agents. We found that in those galaxies with lighter central black-holes, star formation lasted longer. This time delay, consistent with the observed differences in the [$\alpha$-elements$/$Fe] of over- and under-massive BH galaxies\cite{MN16}, is naturally explained if quenching is driven by active galactic nucleus (AGN) feedback. Accretion onto more massive black holes leads to more energetic AGN feedback which would quench the star formation faster. This high-redshift picture has its $z\sim0$ counterpart, as recent star formation is also expected to be regulated by AGN activity. If the energy injection rate scales with the mass of the black hole\cite{Crain15,Sijacki15}, lighter black holes, growing at low accretion rates in the nearby Universe, would be less efficient at keeping the gaseous corona hot, which will ultimately cool and form new stars\cite{Terrazas16}. In Figure~\ref{fig:sfr}, the fraction of young stars anti-correlates with the relative mass of the black hole, further supporting an active role of black holes in regulating star formation within massive galaxies.

Investigating the connection between star formation and black hole activity has been one of the biggest observational challenges since AGNs were proposed as the main source of feedback within massive galaxies. Whereas star formation takes places over longer periods of time, the rapid and non-linear response of black holes to gas accretion\cite{Bower17} complicates a clean empirical comparison of both quantities. AGNs typically populate star forming galaxies, but their luminosities may not correlate with observed star formation rates\cite{Kauffmann03,Mullaney12,Stanley15,RD17}. Here, we made use of the relation between black hole mass and star formation histories to show that the evolution of star formation in massive galaxies over cosmic time is driven by black hole activity. Our results suggest a causal origin for the observed scaling relations between galaxy properties and black hole mass, offering observational support for AGN-based quenching mechanisms.

\bibliography{SFH}

\begin{addendum}
 \item[Acknowledgements] We acknowledge support from the NSF grants AST-1616598, AST-1616710, Marie Curie Global Fellowship, SFB 881 The Milky Way System (subproject A7 and A8) funded by the German Research Foundation, and from grants AYA2016-77237-C3-1-P and AYA2014-56795-P from the Spanish Ministry of Economy and Competitiveness (MINECO). GvdV acknowledges funding from the European Research Council (ERC) under the European Union's Horizon 2020 research and innovation programme under grant agreement No 724857 (Consolidator Grant ArcheoDyn). AJR was supported as a Research Corporation for Science Advancement Cottrell Scholar. IMN would like to thank the TRACES group and Mar Mezcua for their insightful comments, Remco van den Bosch for making the HETMGS data publicly available, and Alina Boecker for her help.
\item[Author Contributions] IMN derived the star formation histories along with TRL, and wrote the text. JPB, AJR, TRL and GvdV contributed to the interpretation and analysis of the results.
\item[Author Information] Correspondence and requests for materials
  should be addressed to IMN (imartinn@ucsc.edu).
\end{addendum}

\begin{center}
{\bf METHODS}
\end{center}

\section{Data quality and spectral fitting}

\
\vspace{0.05in}

\noindent Representative spectra of a low-$\sigma$ and a high-$\sigma$ galaxy are shown in Extended Data Figure~\ref{fig:spectra}, with the best-fitting STECKMAP  model overplotted. Due to the high signal-to-noise of the data (typically above $\sim100$ per \AA), the residuals are typically below one percent ($\sim$0.06 and $\sim$0.08 for these low-$\sigma$ and high-$\sigma$ objects, respectively). 

\vspace{0.05in}

\subsection{Velocity dispersion -- metallicity degeneracy.}
There is a well-known degeneracy between $\sigma$ and galaxy metallicity\cite{Koleva08}, which could potentially affect our measurements of the SFH. It has been shown that fitting kinematics and stellar population properties independently minimizes the effect of this degeneracy\cite{Pat11}. Thus, we first determined the kinematics (V$_\mathrm{sys}$ and $\sigma$) using the penalized pixel-fitting method (pPXF)\cite{ppxf}, which was also used to remove the nebular emission from our spectra. The temporal combination of single stellar population models, convolved to the resolution of the galaxy measured with pPXF, was used to calculate the SFHs. 

As a further test to assess the dependence of our results on the adopted velocity dispersion, we have repeated the analysis while allowing STECKMAP to measure the kinematics at the same time as the SFHs, although this approach has been proved to be  less accurate\cite{Pat11}. In Extended Data Figure~\ref{fig:sigma} we show the cumulative mass distributions of under-massive and over-massive black hole galaxies measured in this way. The observed differences in the SFHs across the black hole mass -- $\sigma$ relation are not due to degeneracies between stellar populations and kinematical properties.

\section{Robustness of the results}

\
\vspace{0.05in}

\subsection{Regularization parameters.}
To assess the robustness of our results, we have varied the two main {\it free} parameters in our analysis. On the one hand, STECKMAP allows for a regularization in both the SFH ($\mu_x$) and the age--metallicity relation ($\mu_Z$) of the different stellar populations models. Effectively, this regularization behaves as a gaussian prior\cite{Ocvirk06}. Figure~\ref{fig:sfr} was calculated using $\mu_x=\mu_Z=10$. Although the choice of these parameters mainly depends on the quality and characteristics of the observed spectra, we repeated the analysis but varying each regularization parameter by two orders of magnitude. In Extended Data Figure~\ref{fig:tests} we demonstrate that our choice of the regularization parameters does not affect the main conclusions of this work, but provides the most stable solutions within the explored range of $\mu_x$ and $\mu_Z$ values.

\vspace{0.05in}
\subsection{Sample selection.}

As described in the main text, the final sample consists of every galaxy in the HETMGS survey with good enough spectra to perform our stellar populations analysis. In practice, this means rejecting spectra with very low signal-to-noise ($\lesssim 10$) and strong emission lines. Possible bias related to these selection criteria are detailed discussed below. No additional constraints have been applied. Our best-fitting relation is based on our own determinations of the velocity dispersion of individual galaxies, homogeneously measured at half \re. Galaxies with black hole masses departing from our best-fitting \mbh-$\sigma$ by more than +0.2 or -0.2 dex were classified as over-massive or under-massive black hole galaxies, respectively. With this criterion, the number of over-massive, standard, and under-massive black hole galaxies is rather similar, 25, 24 and 25 objects, respectively. As reported in previous studies of large sample of black hole masses\cite{Beifiori12,vdb16}, there is no morphological dependence of our best-fitting \mbh-$\sigma$ relation. In Extended Data Figure~\ref{fig:concen} we show the distributions for the concentration parameter C$_{28} \equiv 5 \log R_{80}/R_{20}$ for under-massive and over-massive black hole galaxies. Galactocentric distances $R_{80}$ and $R_{20}$ encompass 80\% and 20\% of the total light of the galaxy, respectively, and were measured using elliptical isophotes \cite{vdb16}. C$_{28}$ is a proxy for the light concentration of galaxies and therefore of their morphology. Higher C$_{28}$ correspond to ellipsoids, whereas lower values are associated with diskier galaxies \cite{Shimasaku01,Courteau07}. As expected, C$_{28}$ behaves similarly across the \mbh-$\sigma$ relation, with a median value of 5.5 and 5.6 for over- and under-massive black hole galaxies, respectively. A two-sided Kolmogorov-Smirnov test indicates that the differences between both C$_{28}$ distributions are insignificant (P-value = 0.82). Thus, over-massive black hole galaxies are morphologically indistinguishable from under-massive black hole objects.

We have also investigated whether the assumed best-fitting \mbh--$\sigma$ relation could lead to spurious results. As an extreme test, we recalculated the mean SFHs but instead of using our best-fitting solution, we assumed the one calculated for a much larger sample of black hole masses\cite{vdb16}. This sample includes objects which were not observed by HETMGS or whose spectra were rejected in our analysis due to the poor quality. Specifically, this alternative \mbh--$\sigma$ relation is 
given by

\begin{center}
 $\log M_\bullet = -4.00 + 5.35 \log \sigma$
\end{center}

\noindent
Note that the use of this equation is not consistent with our sample. Velocity dispersions were calculated differently and  over different radial apertures (0.5\re versus 1\re). Thus, adopting this equation to distinguish between over-massive and under-massive black hole galaxies could potentially affect our conclusions. In spite of all this, Extended Data Figure~\ref{fig:tests} shows that the dependence of the SFH on the mass of the central black hole stands, regardless of the implicit \mbh--$\sigma$ relation.

It is worth noting that the zero point and the slope of the  \mbh--$\sigma$ relation depend on the effective velocity dispersion of individual galaxies, and to a lesser extent, on the assumed mass of the black hole. While the latter is unambiguously defined, the stellar velocity dispersion significantly varies within galaxies. However, the relative distinction between over-massive and under-massive black hole galaxies is relatively insensitive to the characteristics of different samples. The majority of objects ($\sim85$\%) classified as over-massive or under-massive using our best-fitting relation are also over- and -under massive according to other widely-adopted \mbh--$\sigma$ relations\cite{kormendy13,vdb16}. Thus, our classification and therefore our conclusions do not depend on absolute $\sigma$ values.

Finally, we have also considered the possibility that differences in the internal kinematics/morphology of galaxies could bias our conclusions. In particular, under-massive black hole galaxies may be offset from the average \mbh--$\sigma$ relation due to a higher prevalence of disk-like, rotationally supported structures, which in general tend to show younger populations \cite{Kuntschner10}. Observationally, there is no evidence of a morphological or mass concentration dependency of the \mbh--$\sigma$ relation\cite{Beifiori12}, although it has been claimed that pseudo-bulges may follow a different scaling relation\cite{Hu08,Kormendy11,Beifiori12}, which becomes relevant for velocity dispersions $\sigma < 200$\,\kms. In Extended Data Figure~\ref{fig:hm} we show the SFR as a function of look-back time only for galaxies above this velocity dispersion threshold, where the bulk of the population is dispersion supported. It is clear that the observed differences in the star formation processes between over-massive and under-massive black hole galaxies are not due to a morphological or kinematical effect.

\vspace{0.05in}
\subsection{Nebular emission and young stellar populations.}
Although the age sensitivity of the spectra is widely spread over the whole wavelength range\cite{Ocvirk06}, our strongest age sensitive feature is the H$\beta$ line. Thus, we decided to be conservative and remove from the analysis objects with strong emission lines ( Amplitude-over-Noise$>4$), which effectively leads to the old stellar populations shown in Figure~\ref{fig:sfr}. If we include galaxies with stronger emission lines, STECKMAP recovers the expected contribution of younger stars, without softening the differences between over- and under-massive black hole galaxies, as shown in Extended Data Figure~\ref{fig:young}.

\vspace{0.05in}
\subsection{Stellar populations synthesis models.}
The main source of systematic uncertainties in any stellar populations analysis is the choice of a given set of models. We have used the MILES stellar populations synthesis models as a reference, as they provide fully empirical spectroscopic predictions over the explored wavelength range and are best suited for intermediate-to-old stellar populations\cite{miles}. However, given our relatively narrow wavelength coverage, we have also addressed the impact of the stellar populations model choice on our results. We have repeated the analysis of our sample of galaxies but considering three additional sets of models, namely, P\'EGASE-HR\cite{LB04}, the Bruzual \& Charlot 2003\cite{bc03} and the GRANADA/MILES models\cite{GD05,GD10}. As shown in Extended Data Figure~\ref{fig:models}, the differences between over-massive and under-massive black hole galaxies are clear in all three models. The agreement between different models is remarkable given the strong underlying differences among them. As an additional test to the robustness of our analysis, we have also compared our observations to a new set of the MILES models which make use of the BaSTi isochrones \cite{Pietrinferni04}. This final comparison is also included in Extended Data Figure~\ref{fig:models}, further reinforcing the conclusion that our results are not driven by systematics in the stellar populations analysis.

\vspace{0.05in}
\subsection{Galaxy densities and sample biases.}

The statistical properties of galaxies with measured black hole masses do not follow those of the overall population of galaxies. In particular, objects with known black hole masses tend to be denser than the average\cite{Bernardi07,hetmgs}, partially because our ability to measure black hole masses depends on how well we can resolve their spheres of influence \cite{Shankar16}. To test whether or not density is a confounding variable in our analysis, we performed a bi-linear fitting between  \mbh--$\sigma$ and stellar density ($M_\star/R_e^3$). We used publicly available sizes and $K$-band luminosities measurements\cite{vdb16} to obtain a best-fitting relation given by

\begin{equation}
 \log M_\bullet = -1.5 + 4.42 \log \sigma - 0.05 \log M_\star \cdot R_e^{-3} 
\end{equation}

\noindent where $M_\star$ is the stellar mass based on the $K$-band luminosity and $R_e$ the effective radius. This best-fitting relation, consistent with previous studies\cite{Saglia16} (Extended Data Figure~\ref{fig:density}), allowed us to investigate the effect of the black hole at fixed stellar velocity dispersion and density. Notice that this relation is still mostly driven by $\sigma$, and only weakly depends on stellar density. The cumulative mass distribution of over-massive and under-massive black hole galaxies, according to the equation above, is shown in Extended Data Figure~\ref{fig:density}.  We found that, when stellar density is also taken into account, the decoupling between over-massive and under-massive black hole galaxies remains unaltered. Thus, the differences observed in the star formation histories are not due to different galaxy densities. Note also that selection biases in the current sample of black hole masses would, if anything\cite{Gultekin11}, change only the normalization of the \mbh--$\sigma$ relation\cite{Shankar16}. Our approach, by construction, is insensitive to this effect.

Additionally, we have explored possible bias introduced by our rejection criteria, i.e., those galaxies that we did not include in the analysis either because of low signal-to-noise or strong emission features. In Extended Data Figure~\ref{fig:sizesigma} we show the \re -- $\sigma$ distribution for HETMGS galaxies, where the sizes were taken from the 2MASS Extended Source catalog\cite{Jarrett00}, averaged following the method adopted by the ATLAS$^\mathrm{3D}$ team\cite{Cappellari11}. We used the stellar velocity dispersions listed in the HETMGS presentation paper\cite{hetmgs}. The typical sizes of over-massive and under-massive black hole galaxies (R$_e^\mathrm{OM}=2.69\pm0.34$ and  R$_e^\mathrm{UM}=2.49\pm0.37$ kpc, respectively) are consistent with the average population of galaxies with known black hole masses (R$_e^\mathrm{BH}=2.74\pm0.22$ kpc). The same applies to the $K$-band luminosities, with a typical value of $\log L_K^\mathrm{OM}=11.13\pm0.08$, $\log L_K^\mathrm{UM}=11.05\pm0.07$ and $\log L_K^\mathrm{BH}=11.08\pm0.05$ $L_\odot$, for over-massive, under-massive and the complete sample of black hole galaxies, respectively. We therefore conclude that our final sample is not significantly biased compared to the average population of galaxies with measured black hole masses.

\subsection{Code availability.} The STECKMAP code used to derive the SFHs is publicly available at 

\
\url{astro.u-strasbg.fr/~ocvirk/indexsteckmap.html}

\begin{extfig*}
\begin{center}
\includegraphics[width=9cm]{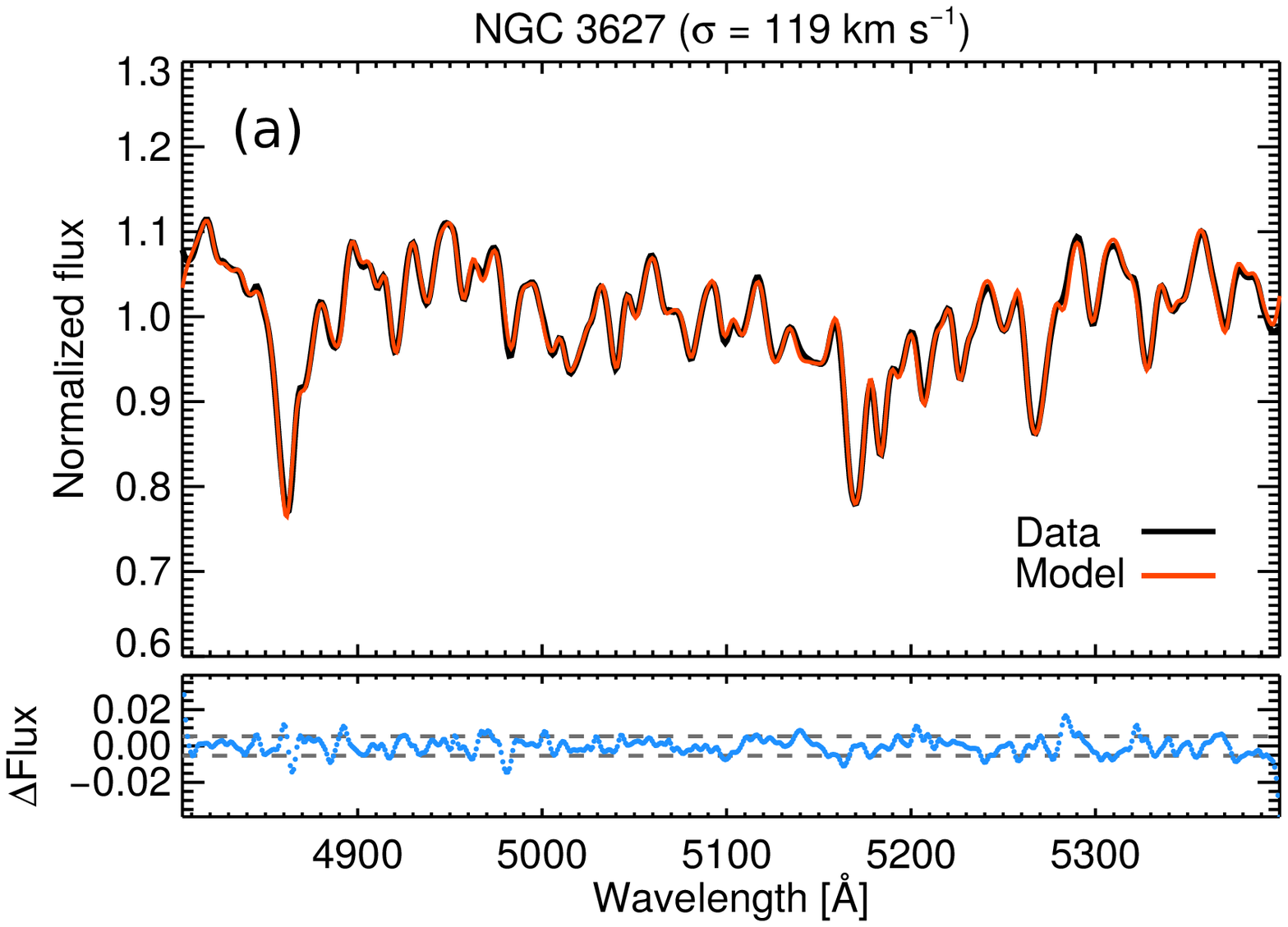}
\includegraphics[width=9cm]{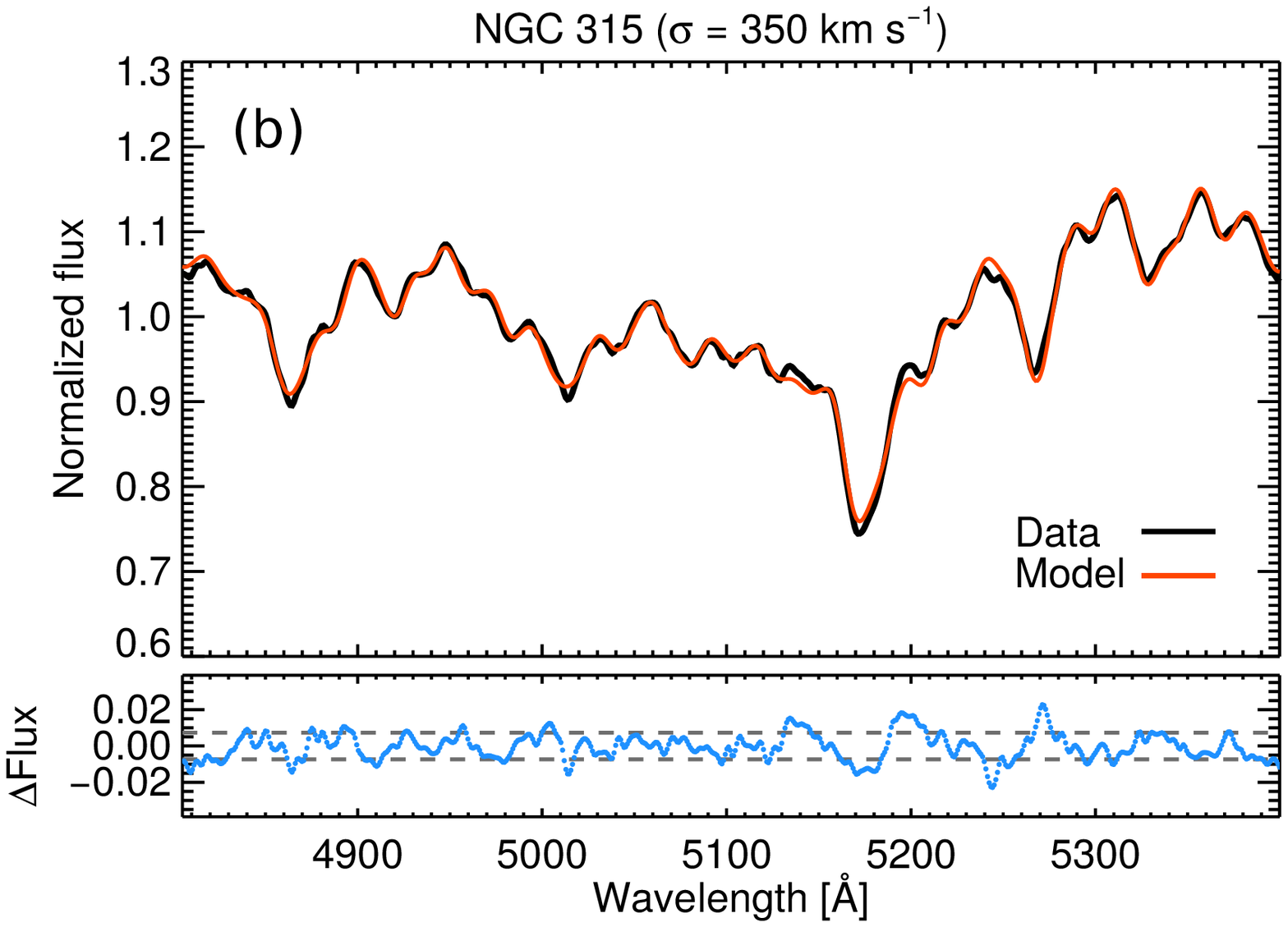}
\end{center}
\caption{{\bf Data and best-fitting stellar populations model.} The spectrum of the low-$\sigma$ galaxy NGC~3627 is shown in panel (a) and the spectrum of the higher $\sigma$ galaxy NGC~315 is shown in panel (b). Along with the HETMGS spectra (black line), we also show the best-fitting STECKMAP model. In the bottom panel we show the residuals (blue dots), which are in both cases below 2\%. The standard deviation is shown as dashed horizontal lines.}
\label{fig:spectra}
\end{extfig*}

\begin{extfig*}
\begin{center}
\includegraphics[width=9cm]{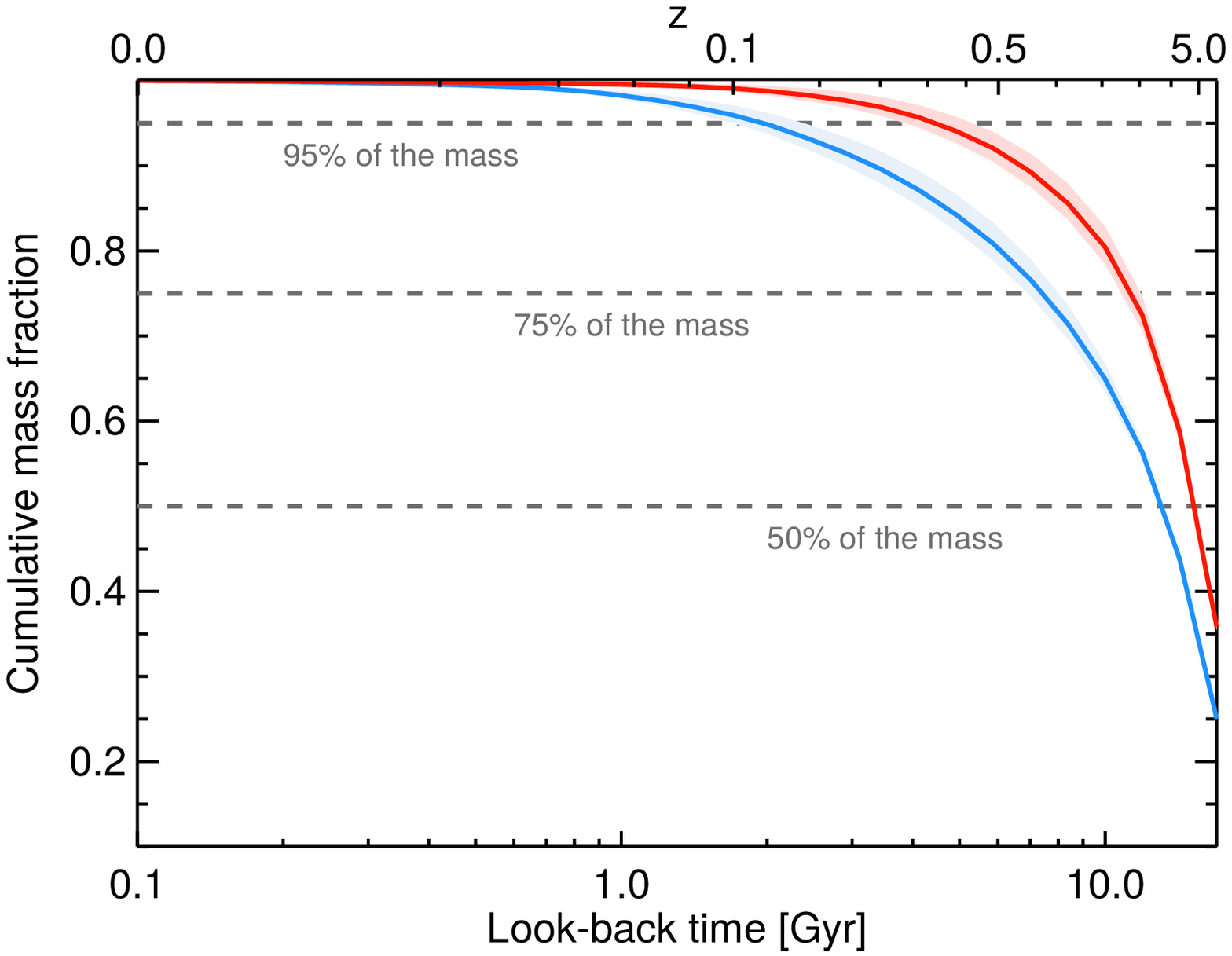}
\end{center}
\caption{{\bf Cumulative mass distributions without fixing the kinematics.} To assess the degeneracy between stellar populations properties and stellar velocity dispersion, we repeated the analysis while allowing STECKMAP to fit the kinematics simultaneously with the SFHs. In red and blue, the cumulative mass fractions of under-massive (blue) and over-massive (red) black hole galaxies are shown as a function of look-back time. Although leaving the kinematics as free parameters leads to less accurate SFHs\cite{Pat11}, the differences in the star formation of under- and over-massive black hole galaxies remain clear.}
\label{fig:sigma}
\end{extfig*}

\begin{extfig*}
\begin{center}
\includegraphics[width=8cm]{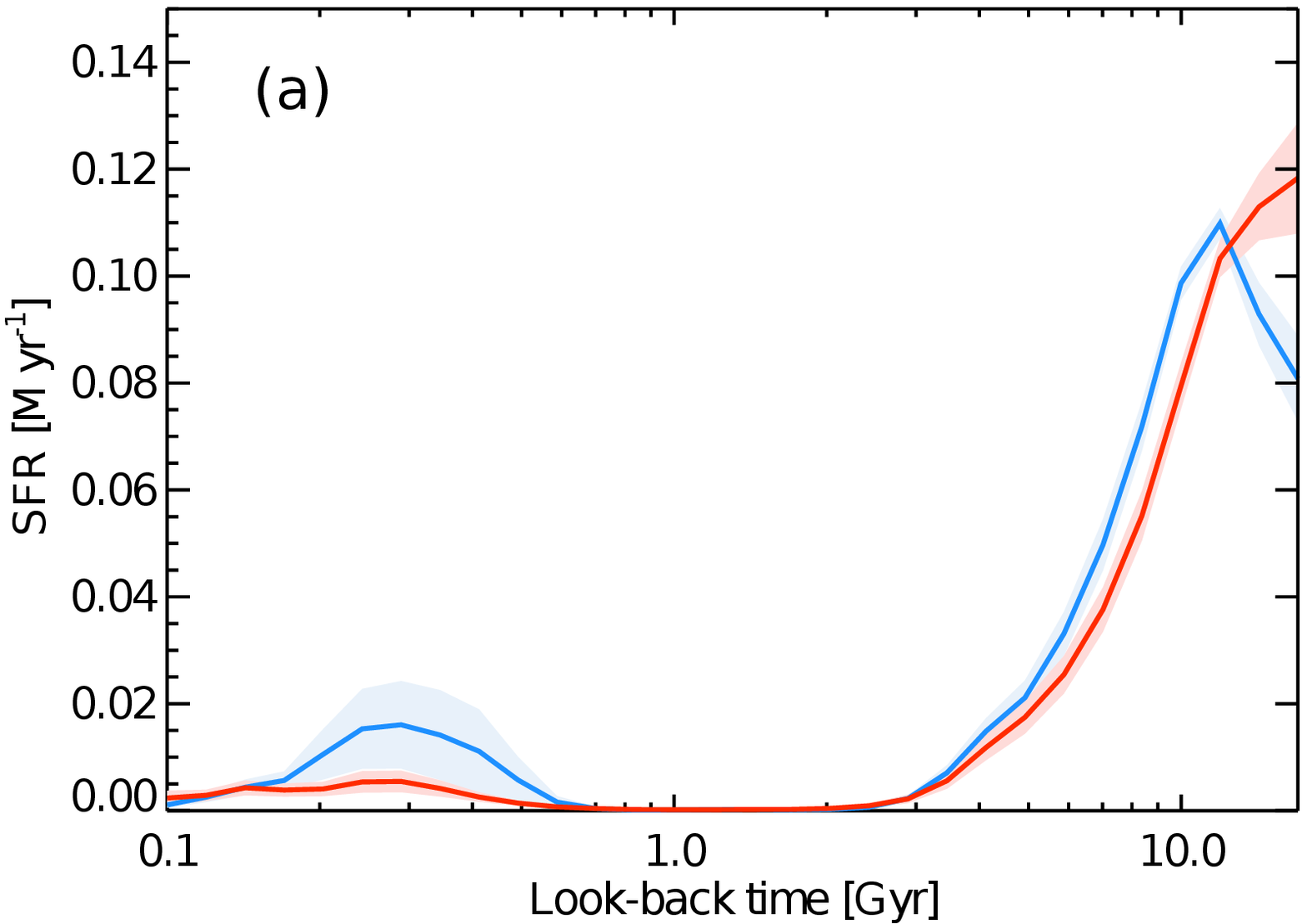}
\includegraphics[width=8cm]{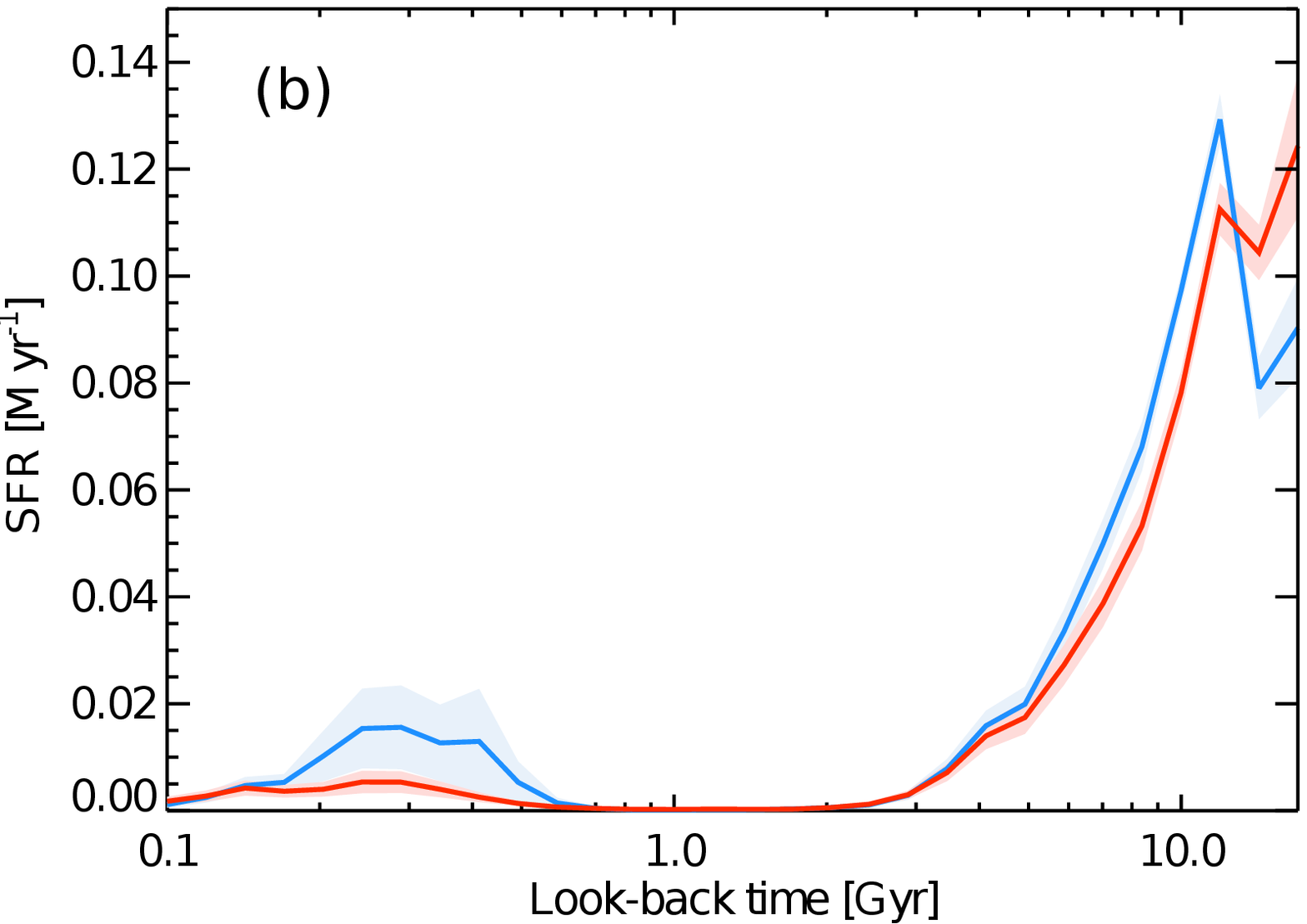}
\includegraphics[width=8cm]{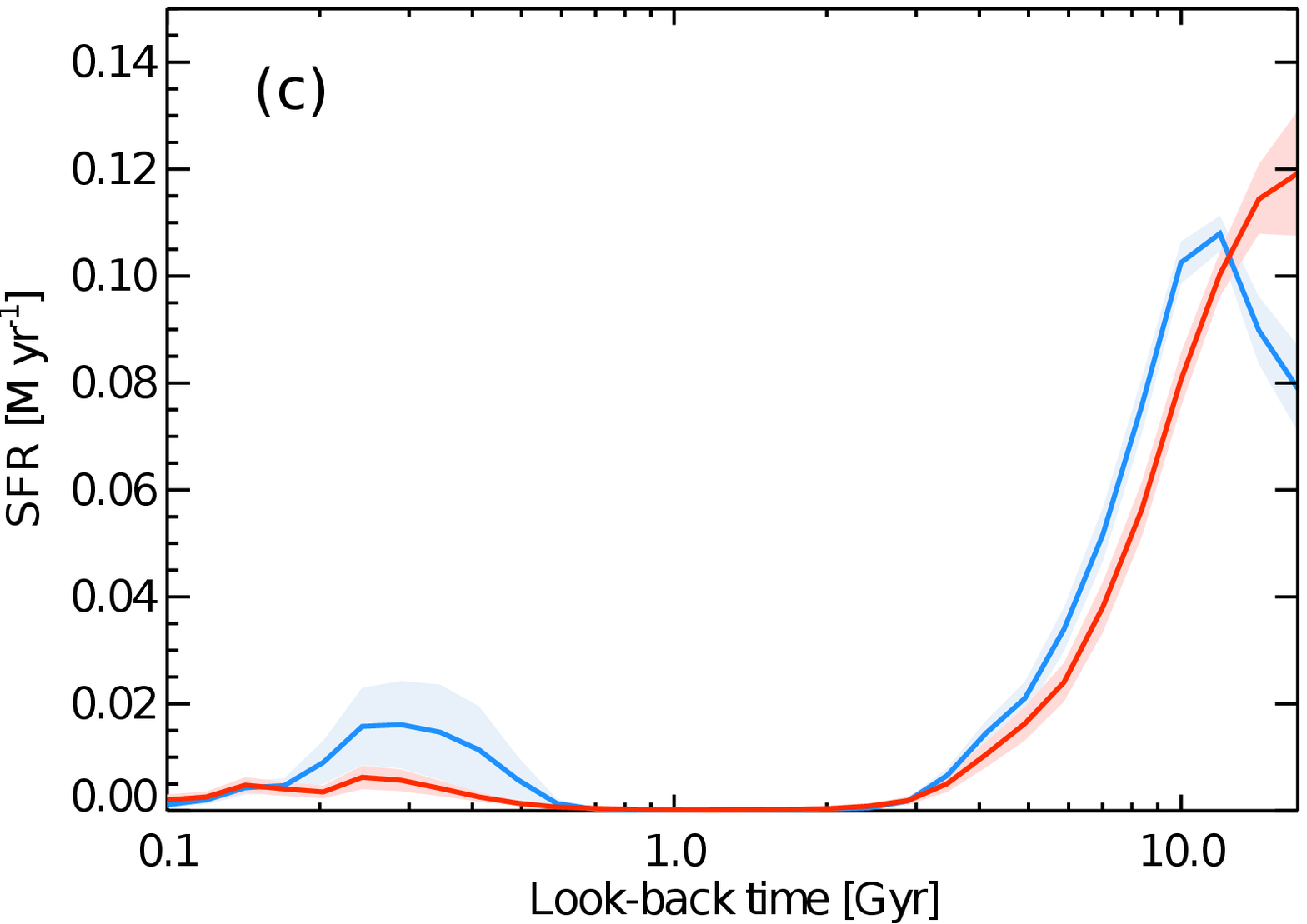}
\includegraphics[width=8cm]{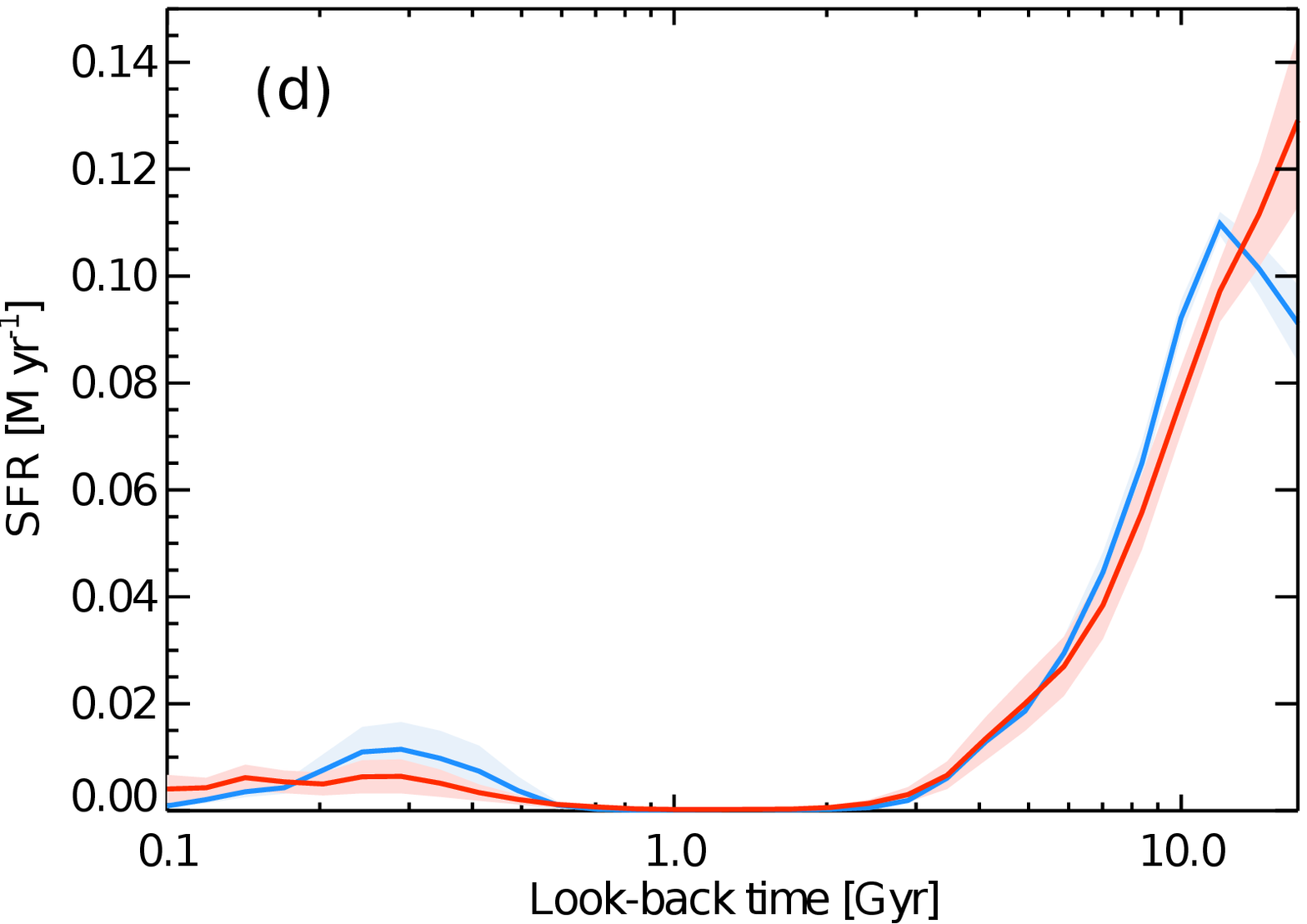}
\end{center}
\caption{{\bf Robustness of the recovered SFRs.} Different panels correspond to the different tests performed in order to explore the reliability of our results. As in Figure~\ref{fig:sfr}, red and blue lines indicate the SFR as a function of look-back time for over- and under-massive black hole galaxies, and the shaded areas mark the 1$\sigma$ uncertainties. As a reference, panel (a) is our preferred model.  A two-sided Kolmogorov-Smirnov comparison between over-massive and under-massive galaxies indicates that both distributions are significantly different (P-value = 0.026). In panel (b) we left the regularization of the SFH almost free, by setting $\mu_x=0.1$. We varied in the same way the regularization of the age--metallicity relation in panel (c), changing $\mu_Z$ from 10 to 0.1. Finally, in panel (d) we adopted a different (but inconsistent) \mbh--$\sigma$ relation\cite{vdb16} to separate our sample. All these tests demonstrate that our conclusions are insensitive to systematics in the analysis.}
\label{fig:tests}
\end{extfig*}

\begin{extfig*}
\begin{center}
\includegraphics[width=9cm]{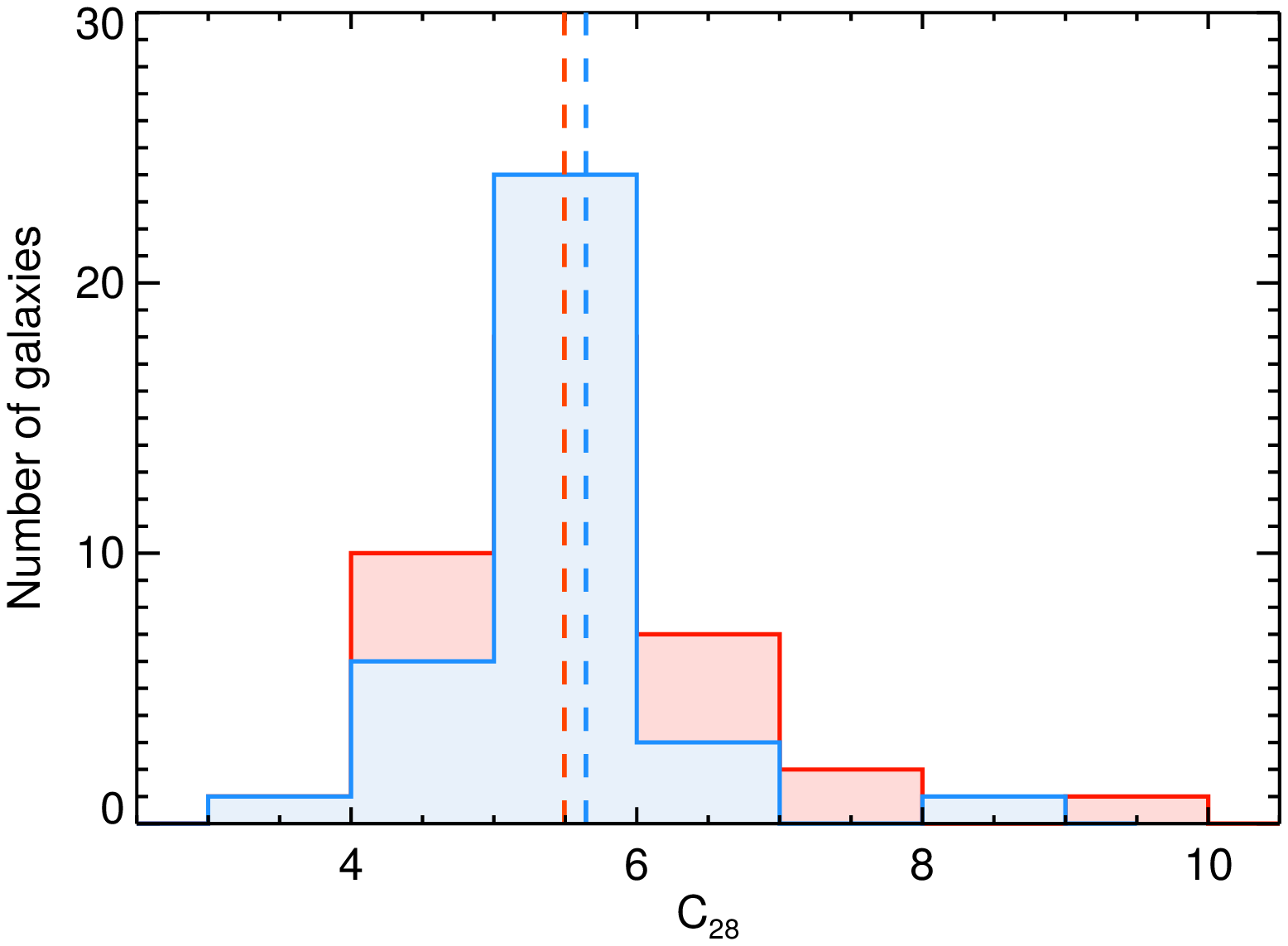}
\end{center}
\caption{{\bf Distributions of concentration parameters.} The distribution of the C$_{28}$ concentration parameter is shown in red and blue for over-massive and under-massive black hole galaxies, respectively. Higher (lower) concentration indices are associated with earlier (later) morphological types. Vertical dashed lines mark the median of the distributions. No significant differences are found between the samples, indicating that the over-massive and under-massive black hole galaxies are morphologically indistinguishable. This suggests that both types of galaxies share the same formation processes, but differ in the present-day mass of their central black holes.}
\label{fig:concen}
\end{extfig*}

\begin{extfig*}
\begin{center}
\includegraphics[width=9cm]{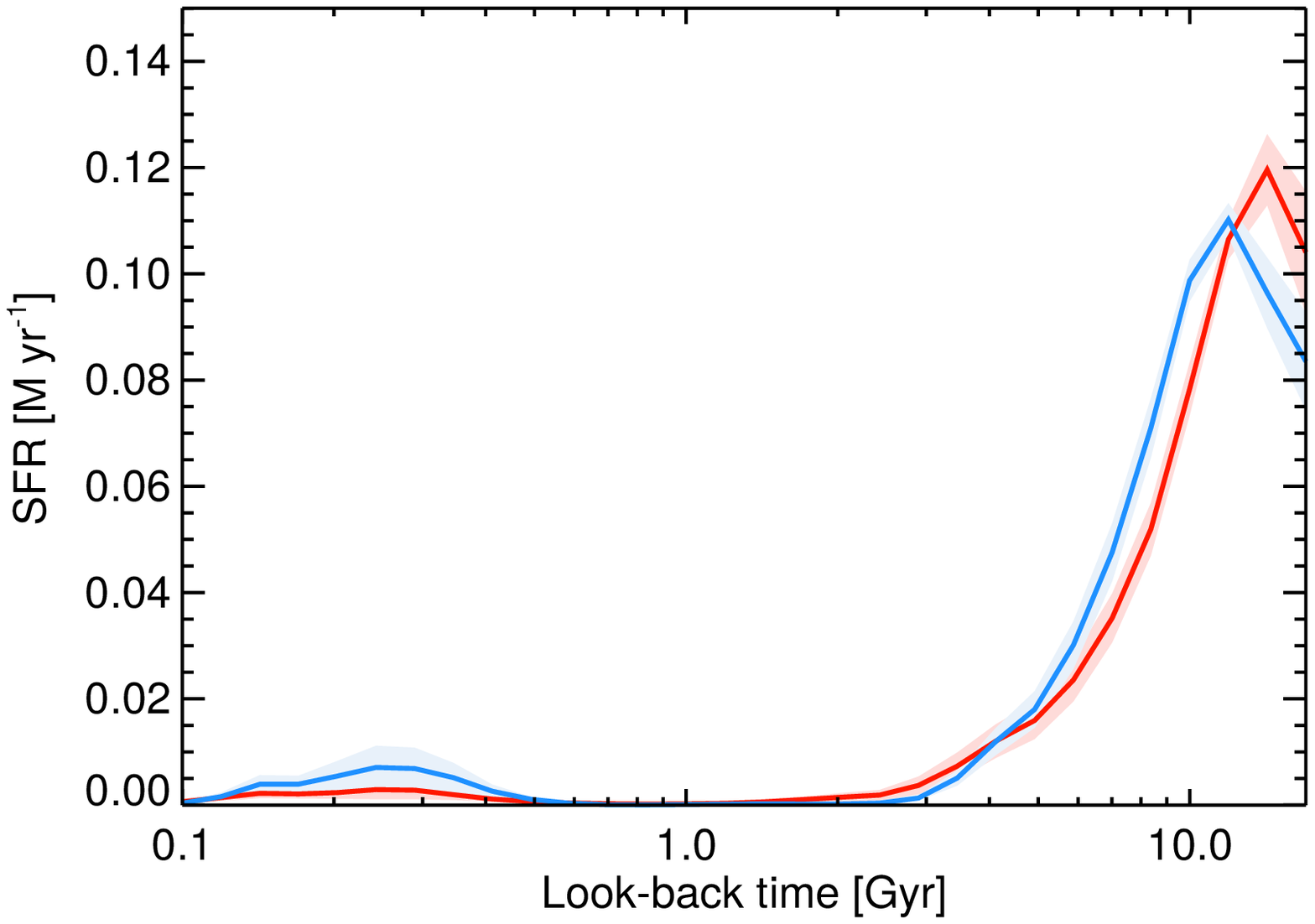}
\end{center}
\caption{{\bf Star formation rate for high-mass galaxies.} Differences in the star formation rate as a function of look-back time for over-massive (red) and under-massive (blue) black hole galaxies, but only including objects with velocity dispersions $\sigma > 200$ \kms ($\log \sigma = 2.3$). The massive end of the \mbh-$\sigma$ relation is dominated by elliptical, pressure-supported systems\cite{Beifiori12,vdb16}, but shows the same distinction between over- and under-massive black hole galaxies.}
\label{fig:hm}
\end{extfig*}

\begin{extfig*}
\begin{center}
\includegraphics[width=9cm]{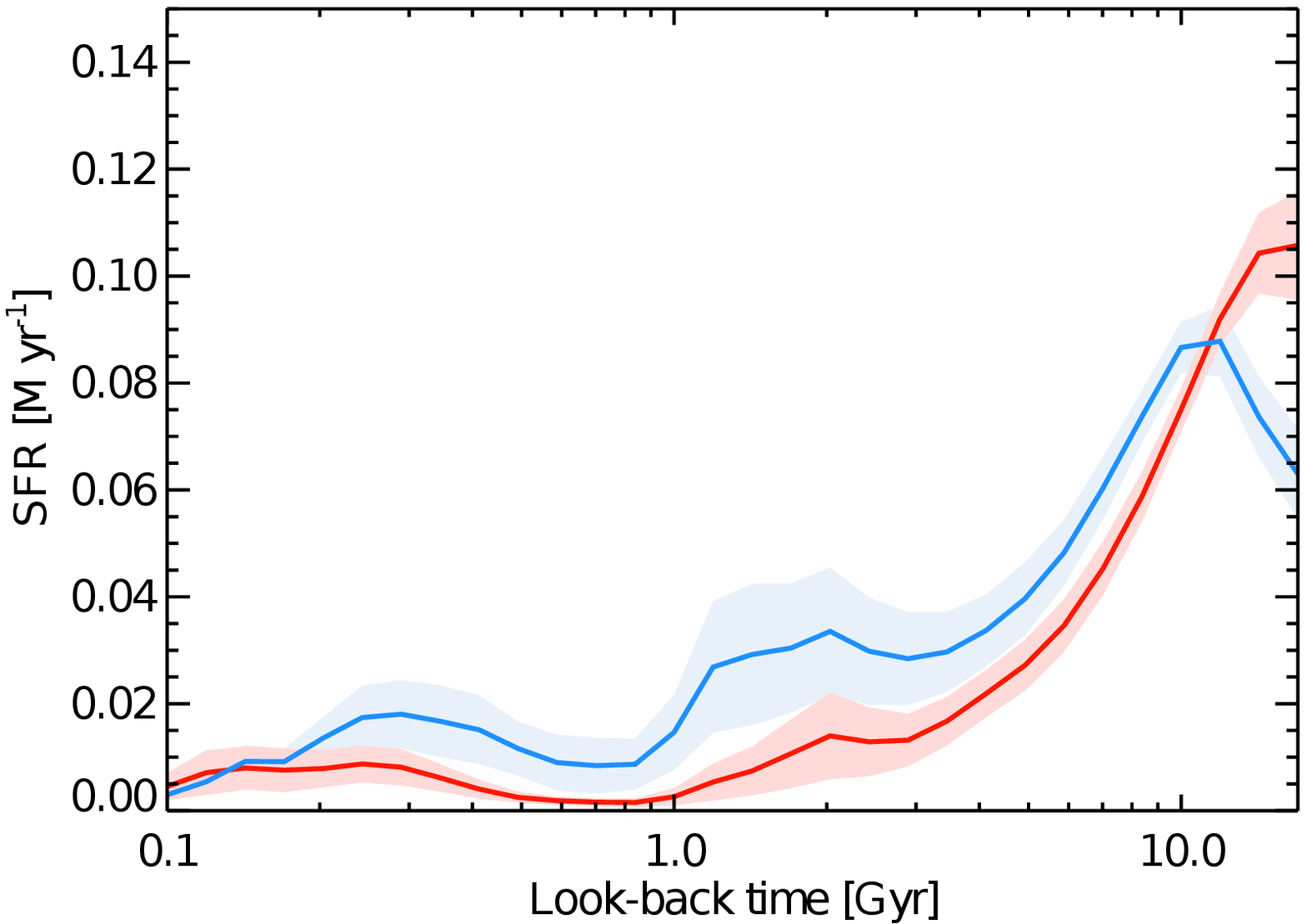}
\end{center}
\caption{{\bf Nebular emission and star formation rate.} If we include in the analysis galaxies with strong emission lines, but the quality of the fits worsens, over-massive (red) and under-massive (blue) black hole galaxies still show decoupled SFHs.}
\label{fig:young}
\end{extfig*}

\begin{extfig*}
\begin{center}
\includegraphics[width=8cm]{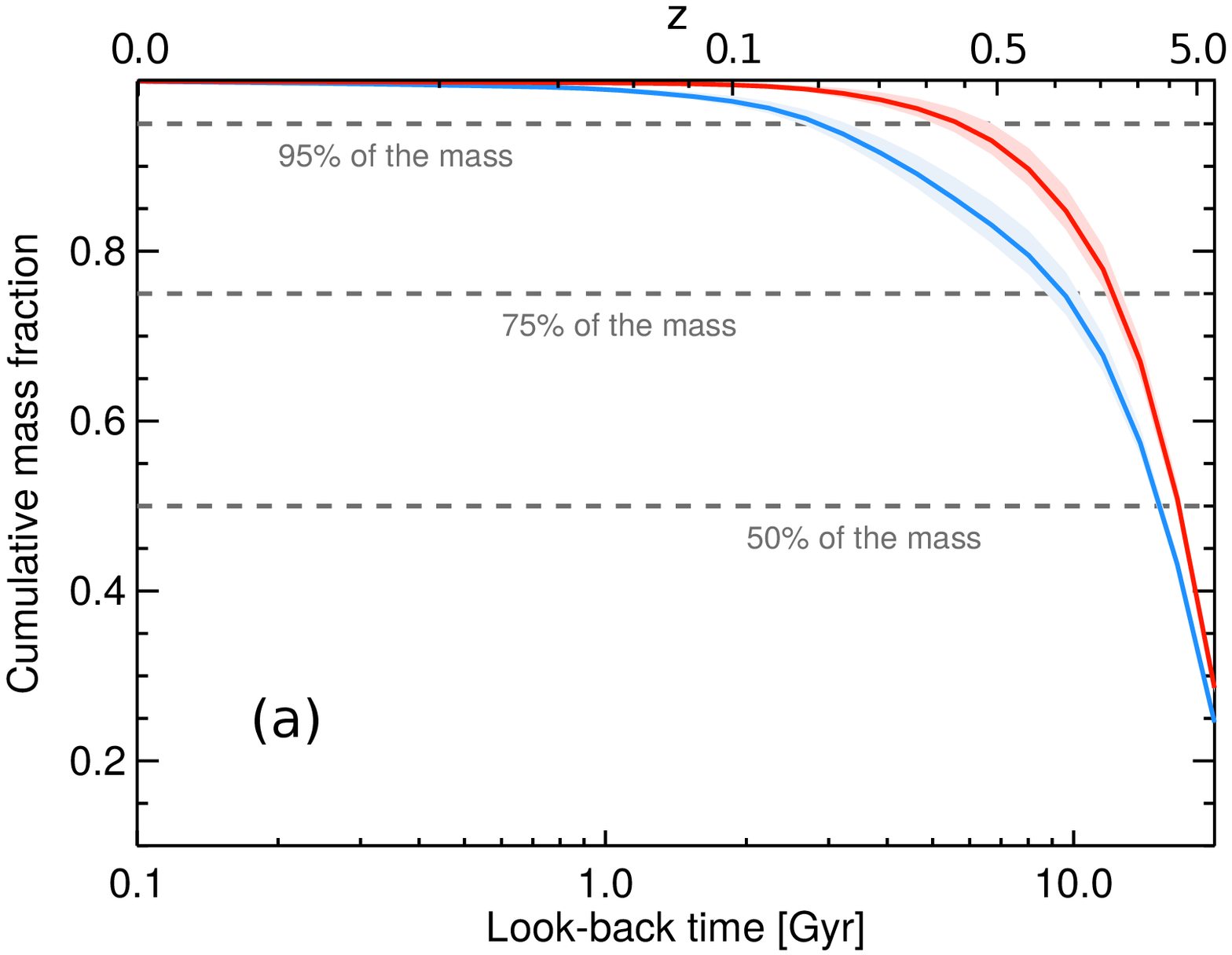}
\includegraphics[width=8cm]{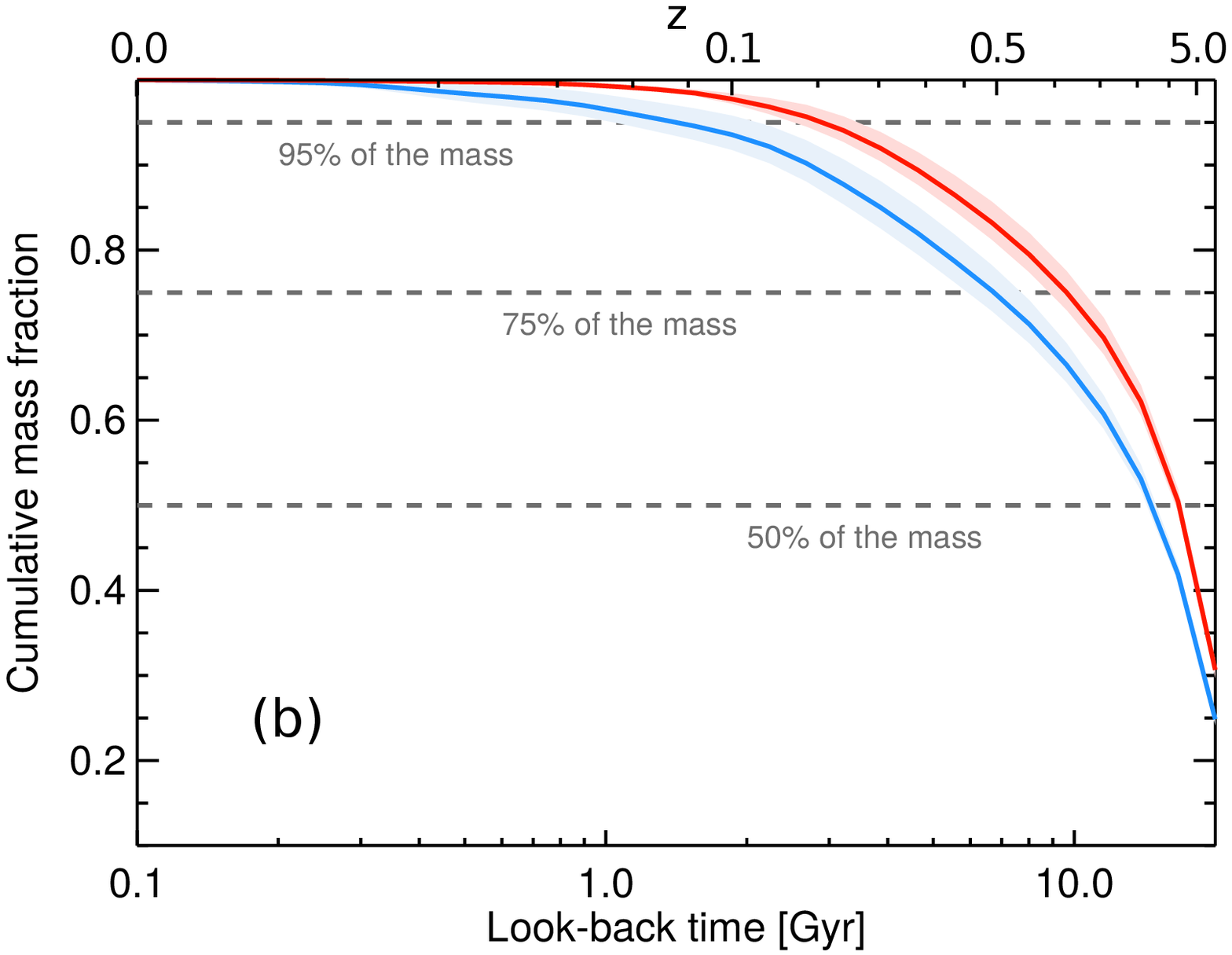}
\includegraphics[width=8cm]{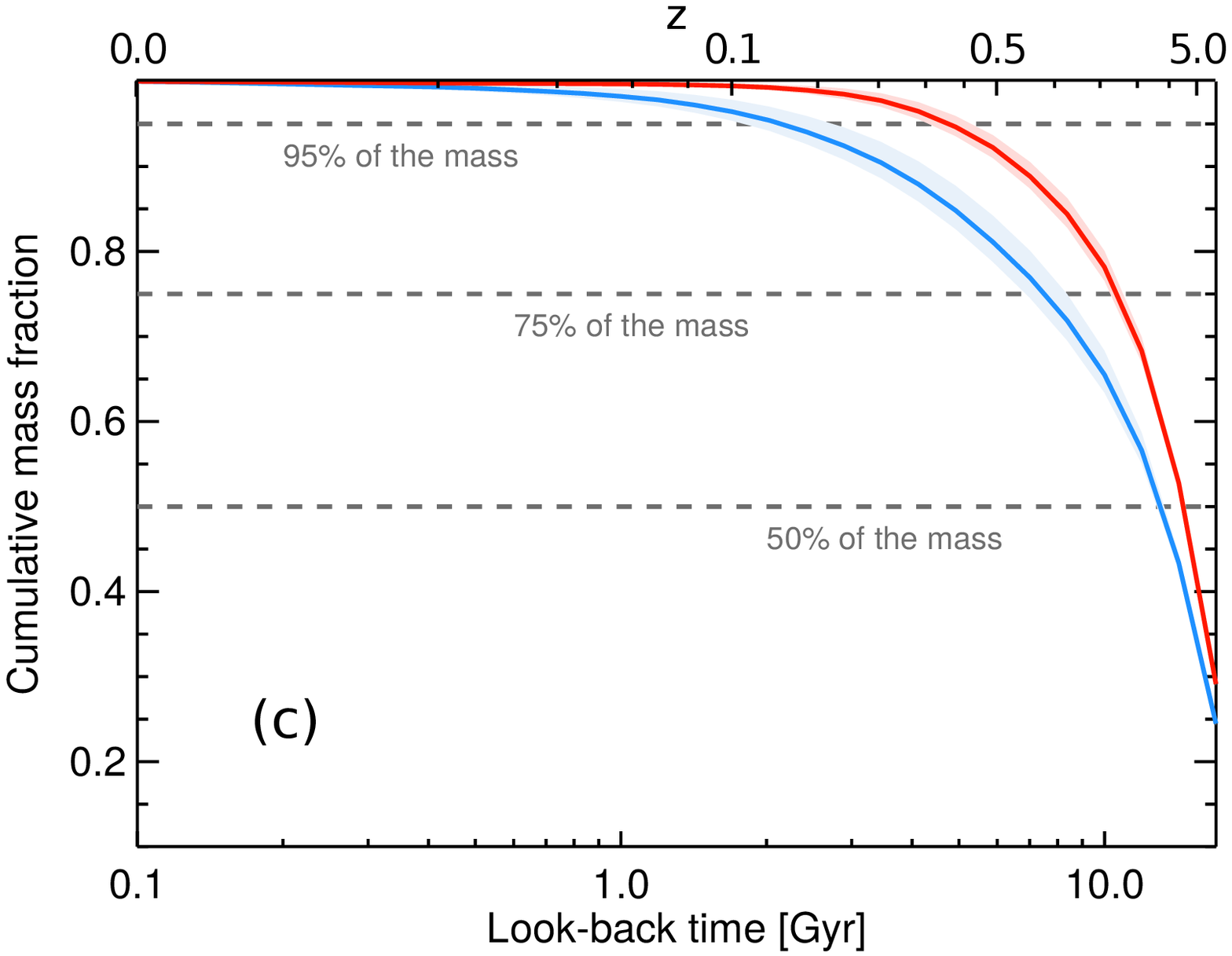}
\includegraphics[width=8cm]{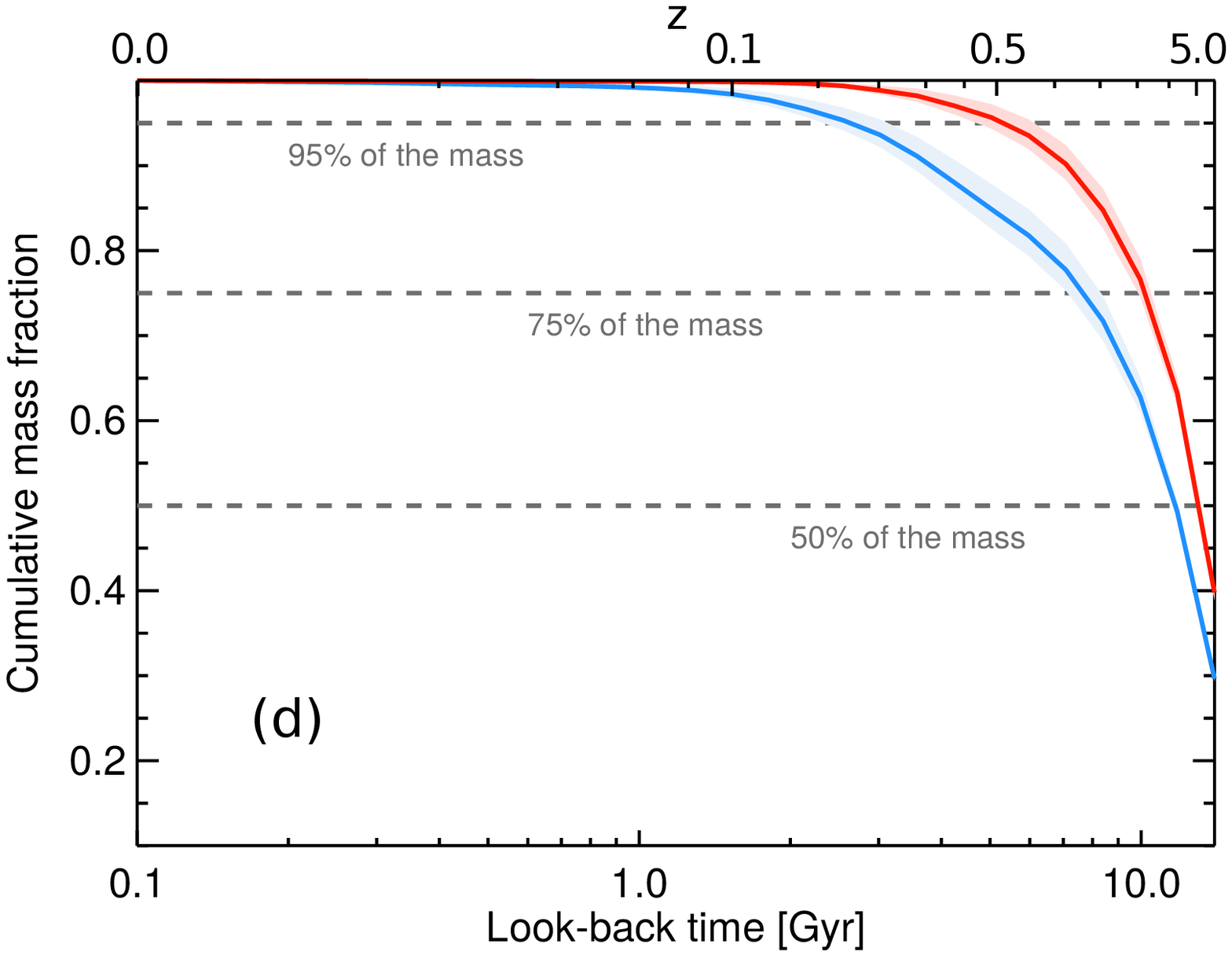}
\end{center}
\caption{{\bf Dependence on the stellar populations models}. As in Figure~\ref{fig:sfr}, each panel shows the cumulative mass fraction for over-massive (red) and under-massive (blue) black hole galaxies. In each panel, the SFHs were calculated using a different set of stellar populations synthesis models. In panel (a) we show the result based on the P\'EGASE-HR\cite{LB04} models, in panel (b) based on Bruzual \& Charlot 2003\cite{bc03}, and in panel (c) using the GRANADA/MILES models\cite{GD05,GD10}. Finally, in panel (d) we use the MILES models with BaSTi isochrones \cite{Pietrinferni04}. Despite the quantitative differences, the different behavior of under-massive and over-massive black hole galaxies is clear in all panels, and thus it is not an artifact of a particular set of models. Note that different models are fed with different stellar libraries, interpolated in different ways to populate different isochrones. The separation between over- and under-massive black hole galaxies is then independent of the different ingredients in stellar population modeling.}
\label{fig:models}
\end{extfig*}

\begin{extfig*}
\begin{center}
\includegraphics[width=8cm]{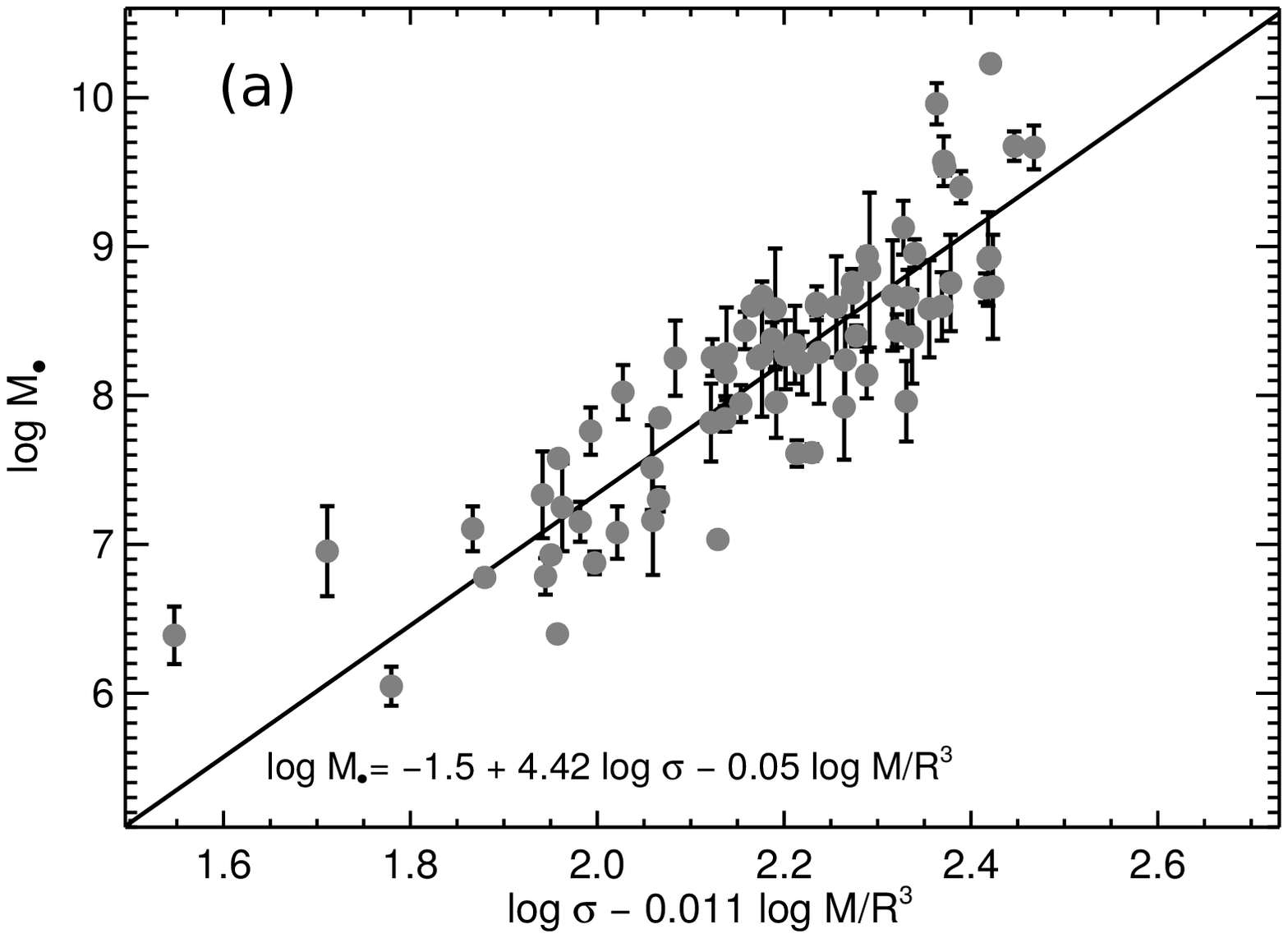}
\includegraphics[width=8cm]{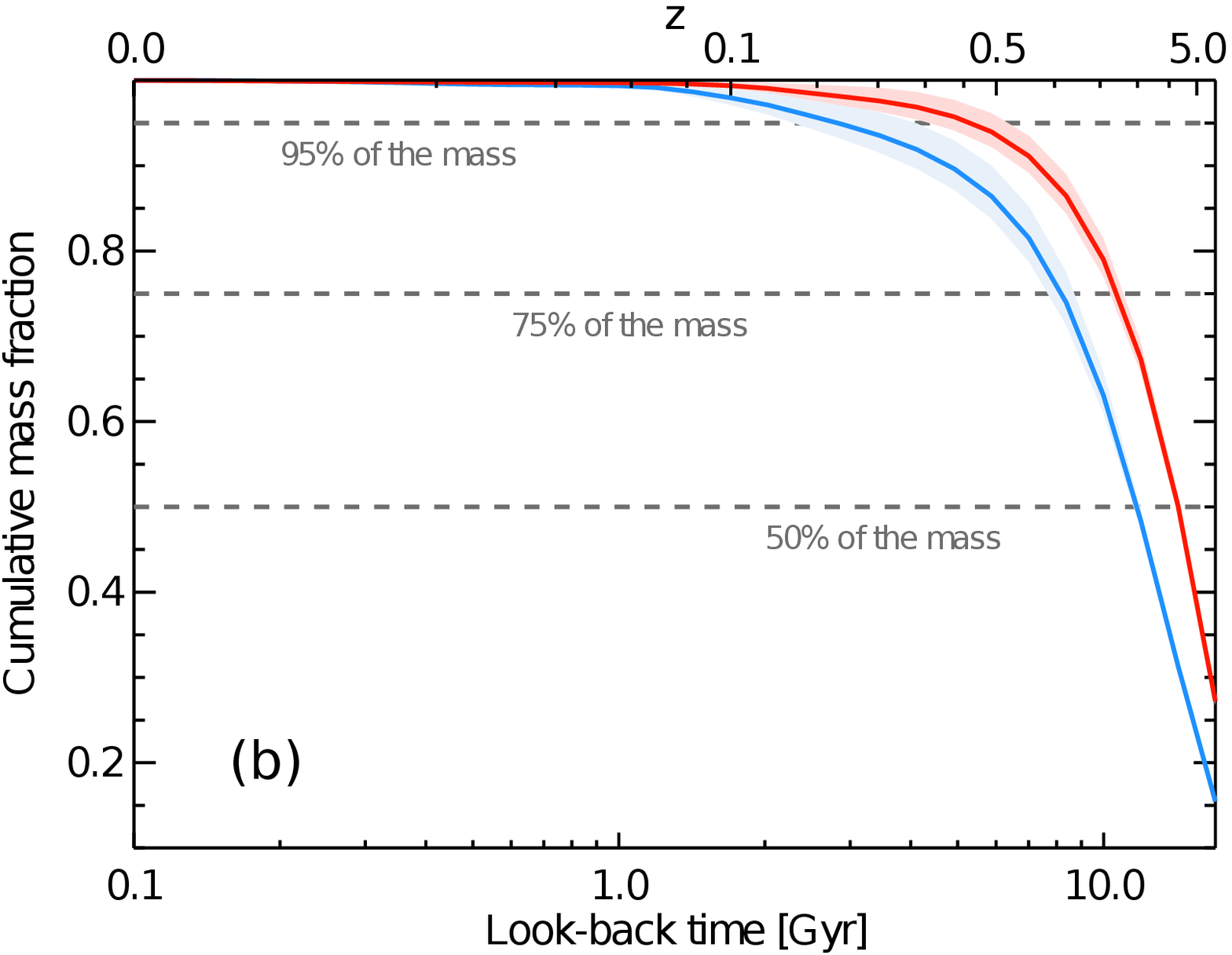}
\end{center}
\caption{{\bf Confounding variables: stellar density.} Panel (a) shows \mbh \ as a function of the best-fitting combination of $\sigma$ and stellar density, following equation~1. Projecting over this plane allows us to separate galaxies depending on the mass of their black holes, but at fixed stellar velocity dispersion and stellar density ($M_\star/R_e^{3}$). The cumulative mass distributions of over-massive and under-massive black hole galaxies according to this new definition are shown in panel (b). This test demonstrates that the observed differences in the SFHs are not due to a possible variation of the stellar density across the \mbh-$\sigma$ relation, further supporting the role of the black hole in regulating star formation within massive galaxies.}
\label{fig:density}
\end{extfig*}

\begin{extfig*}
\begin{center}
\includegraphics[width=8cm]{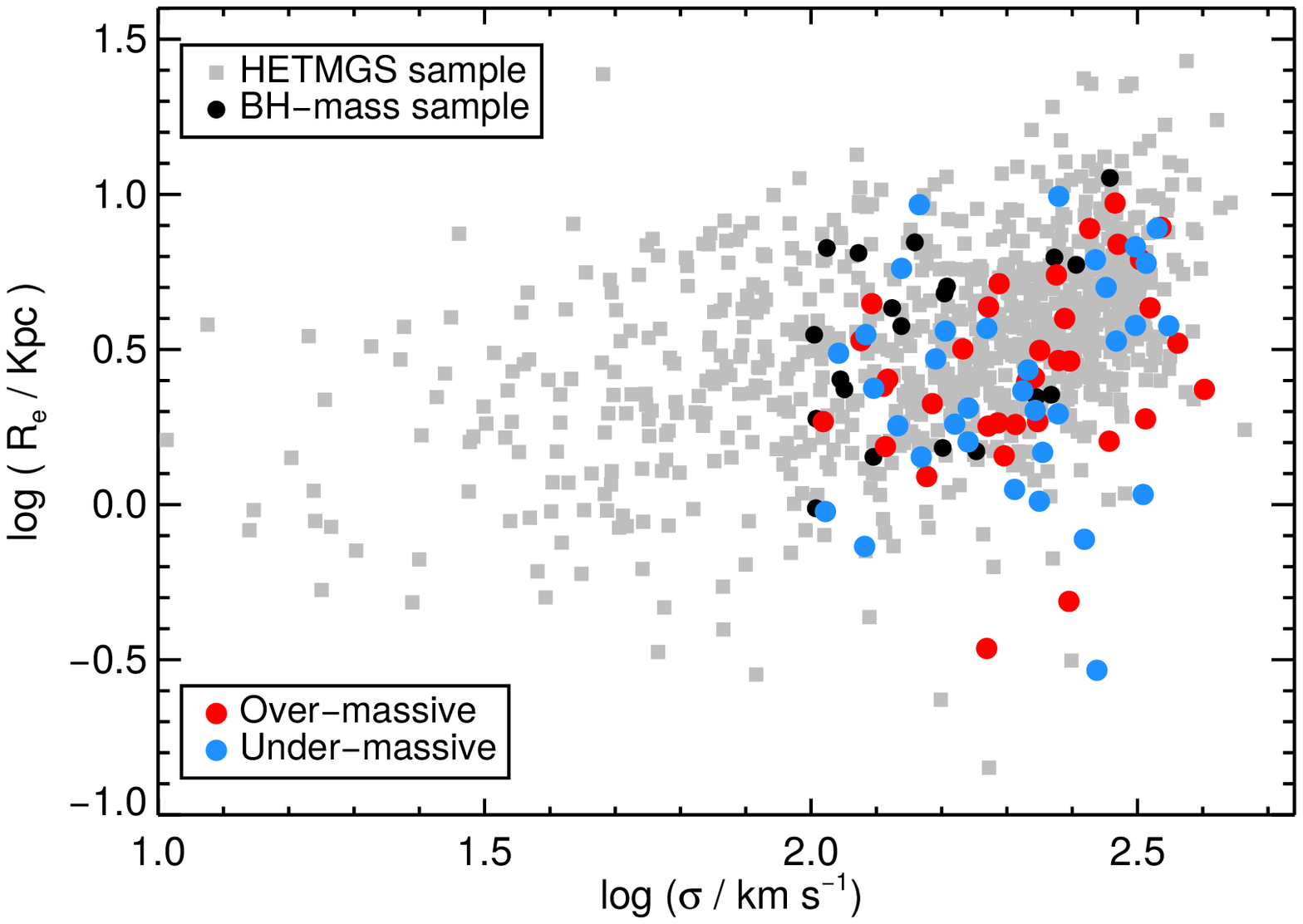}
\end{center}
\caption{{\bf Selection function.} Filled grey squares show the size-$\sigma$ relation for the HETMGS sample, and filled circles correspond to those galaxies with measured black hole masses (blue and red indicate under-massive and overmassive, respectively), which are, on average, more compact than the overall population. However, we found no differences in the mean sizes of our final sample of galaxies (both over-massive and under-massive), compared to the total population of galaxies with known black hole masses (black circles). Thus, our results are not driven by our rejection criteria in terms of signal-to-noise or emission line contamination.}
\label{fig:sizesigma}
\end{extfig*}

\end{document}